\documentclass[aps,prl,twocolumn,floatfix,superscriptaddress,showpacs,longbibliography]{revtex4-1}
\usepackage{graphicx}
\usepackage{amsmath}
\usepackage{amssymb}
\usepackage{amsfonts}
\usepackage{bm}        
\usepackage{dsfont}
\usepackage[mathscr]{eucal}
\usepackage[colorlinks, linkcolor=OrangeRed,citecolor=RoyalBlue,urlcolor=NavyBlue]{hyperref}
\usepackage[normalem]{ulem}
\usepackage[all]{hypcap}
\usepackage[dvipsnames,usenames,table]{xcolor}
\usepackage{siunitx}

\newcommand{\pd}[2]{\frac{\partial #1}{\partial #2}}

\def \Tr{\textrm{Tr}}

\def\be{\begin{equation}}
\def\ee{\end{equation}}

\def\bs#1{\boldsymbol{#1}}

\begin{document}
\title{Dynamical piezoelectric and magnetopiezoelectric effects in polar metals from Berry phases and orbital moments}
\author{D{\'a}niel Varjas}
\affiliation{Department of Physics, University of California, Berkeley, CA 94720, USA}
\affiliation{QuTech and Kavli Institute of Nanoscience, Delft University of Technology, 2600 GA Delft, The Netherlands}
\author{Adolfo G. Grushin}
\affiliation{Department of Physics, University of California, Berkeley, CA 94720, USA}
\author{Roni Ilan}
\affiliation{Department of Physics, University of California, Berkeley, CA 94720, USA}
\affiliation{Raymond and Beverly Sackler School of Physics and Astronomy, Tel Aviv University, Tel Aviv 69978, Israel}
\author{Joel E. Moore}
\affiliation{Department of Physics, University of California, Berkeley, CA 94720, USA}
\affiliation{Materials Sciences Division, Lawrence Berkeley National Laboratory, Berkeley, CA 94720, USA}

\begin{abstract}
The polarization of a material and its response to applied electric and magnetic fields are key solid-state properties with a long history in insulators, although a satisfactory theory required new concepts such as Berry-phase gauge fields.  In metals, quantities such as static polarization and magnetoelectric $\theta$-term cease to be well-defined. In polar metals there can be analogous dynamical current responses, which we study in a common theoretical framework. We find that current responses to dynamical strain in polar metals depend on both the first and second Chern forms, related to polarization and magnetoelectricity in insulators, as well as the orbital magnetization on the Fermi surface.  We provide realistic estimates that predict that the latter contribution will dominate and investigate the feasibility of experimental detection of this effect. 
\end{abstract}
\maketitle


\paragraph{Introduction}

The importance of Berry phases and other geometrical properties of Bloch wavefunctions was first clearly understood in topological phases such as the integer quantum Hall effect~\cite{Thouless82,ass}.  It rapidly became clear that many physical observables in solids are described by Berry phases even in ordinary insulators with no quantization; the electrical polarization in a crystal can be fully and concisely expressed via the Berry connection of Bloch states~\cite{thouless,ksv}.  Metallic systems present additional challenges: in the oldest example, the anomalous Hall effect~\cite{Nagaosa2010}, there are both Berry curvature ``intrinsic''  contributions and ``extrinsic'' contributions that depend on the details of scattering processes.  Discrete symmetries underlie and restrict the emergence of these responses~\cite{Haldane2004}; the anomalous Hall effect is enabled by the breaking of time-reversal symmetry and is observed in magnetic metals.

The goal of this Letter is to analyze a class of transport effects enabled by the breaking of inversion 
symmetry in metals. 
The study of inversion breaking materials such as ferroelectric insulators with switchable polarization, has revealed several fundamental pieces of solid-state physics and lead to a variety of applications~\cite{DRS05,Rabe2007}. 
These advances have translated into a recent increasing interest in the more elusive {\it polar metals}~\cite{Anderson65,SGW13,rondinelli}. While metals do not have a measurable electrical polarization--any surface charge density would be screened by the bulk conduction electrons--polar metals have a low enough symmetry group to support a static polarization were they insulators.
More precisely, we explain how the Berry curvature and related quantities such as the orbital magnetic moments \cite{Xiao2010} 
result in a piezoelectric and magnetopiezoelectric (MPE) response to time-dependent strain
in polar metals with or without time-reversal symmetry.
 Some of these observables can be viewed as generalizations to metals of Berry curvature properties in insulators such as electrical polarization and the orbital magnetoelectric effect, while others are Fermi-surface (FS) properties and hence specific to metals.  
The effects we discuss have important analogues in the corresponding insulating inversion-broken state, in the same way as the integer quantum Hall effect is connected to the intrinsic anomalous Hall effect~\cite{sundaramniu,dahlhaus}. An additional motivation for the present Letter is the active theoretical discussion of when metals, such as Weyl semimetals~\cite{Turner:2013tf,Hosur2013}, can support a current that is induced by and is parallel to an applied magnetic field (the chiral magnetic effect)~\cite{FKW08,Kharzeev2014,Aji2012,burkovcme,Grushin2012,goswamitewari,VF13,changyang}.  The answer is connected to the low-frequency limit of optical activity and involves the magnetic moment of Bloch electrons at the Fermi level~\cite{ZMS16,MP15}, which raises the question of what other properties of metals might involve such magnetic moments. 

The main results of the present Letter are summarized in equations~(\ref{eq:current})-(\ref{eq:alpha2}) and table~\ref{tab:summ} compiles the symmetry requirements for the effects to emerge. The first is referred to as piezoelectricity~\cite{vanderbiltpiezo,Resta94,Resta2007}; in a polar material, even in a metal, any time-dependent change of the material, such as a time-dependent strain, will induce a current resulting from the change of polarization.  In a metal, only changes in polarization are well-defined as these involve measurable bulk currents through the unit cell.  
As a difference with the insulating case where the energy gap protects against process that do not excite electrons far from the ground state, 
we will require a slow evolution of strain relative to electronic time scales~\footnote{
A different mechanism is considered in~\cite{Cortijo2016} that discusses a strain-induced chiral magnetic effect in Weyl semimetals. Unlike the near-equilibrium or adiabatic physics found here, it depends on a non-equilibrium electronic configuration created by straining faster than the intervalley relaxation time. This rapid strain can create a large current in a magnetic field.}.  This assumption guarantees that the distribution function remains close to equilibrium.  Additional effects from strongly non-equilibrium distributions and the scattering processes that restore equilibrium are left for future work.
\begin{table}[t]
\centering
\label{tab:summ}
\caption{The dynamical current effects considered in this Letter and their requirements in terms of inversion $(\mathcal{I})$ and time-reversal $(\mathcal{T})$ symmetries and the orbital moment $\mathbf{m}$.  ``MPE'' stands for magnetopiezoelectricity, i.e., strain-induced currents linear in applied magnetic field.}
\begin{tabular}{@{}llllc@{}}
\toprule
 &  $\mathcal{I}$ & $\mathcal{T}$ & $\mathbf{m}$ & Eq. \\ 
 \colrule
 Piezoelectricity & No & Any & Any & \eqref{eq:beta} \\
 MPE Fermi sea &  No & No & Any & \eqref{eq:alpha1}\\
 MPE Fermi surface &  No & No & Nonzero & \eqref{eq:alpha2} \\
\botrule
\end{tabular}
\end{table}

\begin{figure}[b]
\begin{center}
\includegraphics[width=8cm]{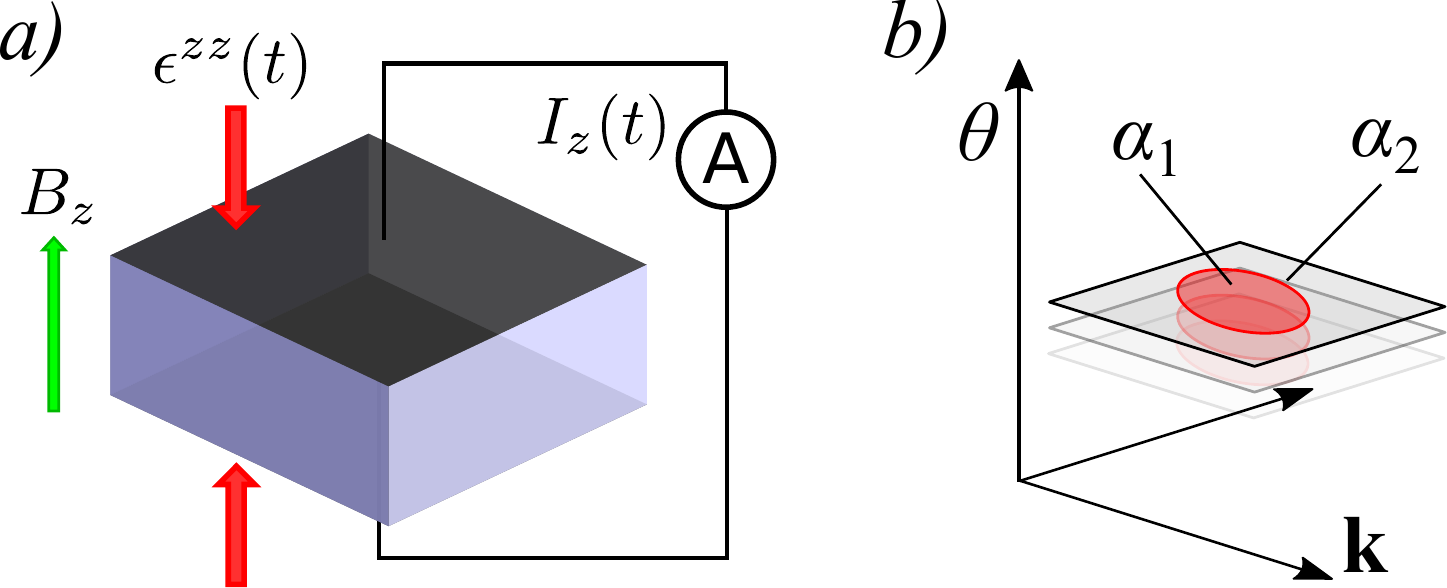}
\caption{a)~Schematic experimental setup. The sample is placed in static magnetic field and homogenous time-dependent strain is appiled. The top and bottom surfaces are contacted, short-circuited through a low impedance ammeter and the current parallel to the applied field is measured. b)~Illustration of the four dimensional momentum space. The time-dependent parameter $\theta$ spans an orthogonal direction to the three dimensional Brillouin zone, projected to two dimensions here. The contributions to the MPE $\alpha_1$ and $\alpha_2$ come from the interior of the Fermi sea (shaded red) and the Fermi surface (red contour) respectively.}\label{fig:MPE}
\end{center}
\end{figure}

A second effect, which we call magnetopiezoelectricity, 
emerges when the material is magneticaly ordered and time-reversal symmetry is broken along with inversion symmetry. This second order current response is bilinear in strain rate and static external magnetic field (Fig.~\ref{fig:MPE}~a). One contribution can be viewed as the generalization to metals of the orbital magnetoelectric effect in insulators~\cite{qilong,essinmoorevanderbilt,Essin2010,malashevich}.
It involves the second Chern form of the Berry gauge fields~\cite{Jian2013,Barkeshli2011}\footnote{Refs.~\cite{Jian2013} and \cite{Barkeshli2011} discuss a different occurrence of the second Chern form in the response of metals to magnetic field in the presence of static spatial inhomogeneity rather than the purely dynamical effects studied here.}, a slightly more complicated geometrical object than the first Chern form that controls the polarization and Hall effect, and can be interpreted as a metallic version of the dynamical axion effect in antiferromagnets~\cite{Li2010,Sekine2016}. 

We also find a second, purely Fermi-surface contribution to the MPE that is proportional to the orbital magnetic moment. Our estimates for realistic systems suggest that this part of the MPE
unique to metals dominates the response. It is therefore the main prediction of this letter for a new experimental effect.
\paragraph{Methodology} 
To address the topological responses of metallic magneto-electrics we employ the semiclassical formalism~\cite{Xiao2007,Xiao2009,sundaramniu,
Xiao2010,Jian2013,Gao2015}.
Our starting point is a three-dimensional Hamiltonian of a metal $H(\mathbf{k},{\theta})$ 
that is parametrized by a time-dependent parameter ${\theta}(t)$.
The microscopic origin of ${{\theta}(t)}$ can be diverse, it can for example parametrize ferromagnetic~\cite{Wang2016} or antiferromagnetic ordering~\cite{Li2010}. Such a fluctuating magnetic order in insulating systems has been previously studied~\cite{Ooguri2012} and termed ``dynamical axion field". 

In this Letter we focus on the case where ${\theta}$ emerges from the coupling of homogeneous time-dependent strain 
to orbital degrees of freedom, which effectively renormalizes the hopping structure of $H(\mathbf{k})$ in a time-dependent fashion
leading to $H(\mathbf{k},{\theta})$. The parameter ${\theta}$ can refer to any strain component, or an arbitrary parametrization of some combination of strain components. 
Before proceeding, it is worth highlighting several relevant aspects of our calculation. First, strain is non-electromagnetic and acts as an independent external field. 
Second, although we allow for the time-reversal-breaking magnetic order required for the MPE to depend on $\theta$, we assume it 
does not respond to external magnetic fields at the linear order of interest here. Thus we only focus on the orbital contribution.
Finally, we assume the clamped ion limit; strain changes the hopping amplitudes for the electrons but the atomic coordinates remain fixed. 

A compact way of dealing with  $H(\mathbf{k},{\theta})$ is to regard ${\theta}$ as an extra momentum coordinate.
The semiclassical equations governing the dynamics in this case are given by
\begin{eqnarray}
\dot{r}^{i} &=& \frac{1}{\hbar} \pd{\tilde{\mathcal{E}}_{\mathbf{k},{\theta}}}{k_{i}}-\left(\tilde{\mathbf{\Omega}}\times \dot{\mathbf{k}}\right)^i- \tilde{\Omega}^{i\theta}\dot{\theta},\\
\hbar\dot{k}_{i} &=& -eE_{i}-  e\left(\mathbf{B}\times \dot{\mathbf{r}}\right)_i,
\end{eqnarray}
in terms of the external magnetic ($\mathbf{B}$) and electric ($\mathbf{E}$) fields, the $i=x,y,z$ component of the three-dimensional position ($\mathbf{r}$) and momentum ($\mathbf{k}$). The Berry curvature components $\tilde{\boldsymbol{\Omega}}$ and $\tilde{\Omega}^{i\theta}$, to be defined precisely below, 
determine the Hall conductivity~\cite{thouless} and the piezoelectric effect~\cite{Resta2007} respectively.
For what follows, we find it convenient to promote the semiclassical picture to a four-dimensional space defined by an extended momentum and position vector (see Fig.~\ref{fig:MPE}~b), $k_{\mu}=(\mathbf{k},{\theta})$ and $\mathbf{r}_{\mu}=(\mathbf{r},r_{{\theta}})$ respectively,
with $\mu=x,y,z,{\theta}$ ~\footnote{We use Latin letters to represent the spatial components $x,y,z$ and Greek letters for the four component vectors  $x,y,z,\theta$}.
The semiclassical equations for such a phase space read~\cite{Xiao2010,PZO15}
\begin{eqnarray}
\label{eq:dotr4D}
\dot{r}^{\mu} &=& \frac{1}{\hbar} \pd{\tilde{\mathcal{E}}_{\mathbf{k},{\theta}}}{k_{\mu}}-\tilde{\Omega}^{\mu\nu}\dot{k}_{\nu},\\
\label{eq:dotk4D}
\hbar\dot{k}_{\mu} &=& -eE_{\mu}-e B_{\mu\nu} \dot{r}^{\nu}.
\end{eqnarray}
Here $E_{\mu}$ and the antisymmetric tensor $B_{\mu\nu}$ are the generalization of the electric $(E_{i})$ and magnetic ($B_{i} = \frac{1}{2}\epsilon_{ijk} B_{jk}$) fields
where by construction $B_{\mu{\theta}}=0$ that implies, from \eqref{eq:dotk4D}, that $\hbar\dot{{\theta}}=-eE_{{\theta}}$.
We note that Eq.~\eqref{eq:dotr4D} includes two corrections due to the external fields~\cite{GYS14, Xiao2010}.
One modifies the band structure 
$\varepsilon_{\mathbf{k},{\theta}}\to \tilde{\mathcal{E}}_{\mathbf{k},{\theta}}=\varepsilon_{\mathbf{k},{\theta}}-\mathbf{m}_{\mathbf{k},{\theta}}\cdot\mathbf{B}$ 
where 
$\mathbf{m}_{\mathbf{k},{\theta}}=\frac{e}{2\hbar}\mathrm{Im} \langle \partial_{\mathbf{k}} u_{\mathbf{k},{\theta}} | \times (H-\varepsilon_{\mathbf{k},{\theta}})|\partial_{\mathbf{k}} u_{\mathbf{k},{\theta}}  \rangle$ 
is the magnetic orbital moment defined through the Bloch wave-functions $| u_{\mathbf{k},{\theta}}  \rangle$.
Second, the unperturbed Berry curvature $\Omega^{\mu \gamma} =  \partial_{k_\mu} a_{k_\gamma }  -  \partial_{k_\gamma} a_{k_\mu}$ where 
$a_{k_\mu} = i \langle u_{\mathbf{k},{\theta}} |\partial_{k_\mu }| u_{\mathbf{k},{\theta}} \rangle $ is corrected as 
$\Omega^{\mu \gamma} \to \tilde{\Omega}^{\mu \gamma} = \Omega^{\mu \gamma} +  \Omega_{1}^{\mu \gamma}$ ($\tilde{\Omega}_{i}=\frac{1}{2}\epsilon_{ijk}\tilde{\Omega}^{jk}$).
The additional $\Omega_1^{\mu \gamma} =  \partial_{k_\mu} a'_{k_\gamma }  -  \partial_{k_\gamma} a'_{k_\mu}$ is defined by 
$a'_{k_\mu } = i \langle u_{\mathbf{k},{\theta}} |  \partial_{k_\mu } | u'_{\mathbf{k},{\theta}} \rangle + \mbox{c.c.}$ that incorporates the first-order correction to the Bloch wave-function $|u'_{\mathbf{k},{\theta}} \rangle$.
The quantity $a'_{k_\mu}$ is gauge-invariant, and physically corresponds to a shift of the wave-packet centre induced by interband mixing from the external fields \cite{GYS14}.

Combining \eqref{eq:dotr4D} and \eqref{eq:dotk4D} and keeping terms to second order in the external fields results in~\cite{PZO15}
\begin{align}
\nonumber
\dot{r}^{\mu} =& \frac{1}{\hbar}\pd{\tilde{\mathcal{E}}_{\mathbf{k},{\theta}}}{k_{\mu}} + 
\frac{e}{\hbar}\tilde{\Omega}^{\mu\nu} 
\left( E_{\nu} + \frac{1}{\hbar} B_{\nu\lambda} \pd{\tilde{\mathcal{E}}_{\mathbf{k},{\theta}}}{k_{\lambda}} \right) \\
&+ \frac{e^2}{\hbar^2}\tilde{\Omega}^{\mu\nu} B_{\nu\lambda}\tilde{\Omega}^{\lambda\gamma}
\left( E_{\gamma} + \frac{1}{\hbar} B_{\gamma\delta} \pd{\tilde{\mathcal{E}}_{\mathbf{k},{\theta}}}{k_{\delta}} \right) +\cdots \, ,
\label{eq:dotronly}
\end{align}
which enters the current density 
\begin{equation}
j^\mu = e\int_{\mathbb{T}^3} { \text{d}^3 k } [ \dot{{ r}}^\mu D_{{\bs k},\theta}]f(\tilde{\mathcal{E}}_{\mathbf{k},{\theta}},\mu). 
\label{eq:jdef} 
\end{equation}
Here $f(\tilde{\mathcal{E}}_{\mathbf{k},{\theta}},\mu)$ is the Fermi-Dirac distribution for the perturbed band structure $\tilde{\mathcal{E}}_{\mathbf{k},{\theta}}$ at chemical potential $\mu$ and $D_{{\bs k},\theta}$ is the modified density of states defined as $D_{{\bs k},\theta} = \left[  1 + \frac{1}{2}\frac{e}{\hbar}  B_{\mu \nu} \tilde{\Omega}^{\mu \nu} + \mathcal{O}(B^2) \right].$
Using \eqref{eq:dotronly} and \eqref{eq:jdef} the current density reads
\begin{align}
\nonumber
 j^\mu =& e \int_{\mathbb{T}^3} \frac{ \text{d}^3 k }{(2\pi)^3} 
  \left[ \frac{1}{\hbar}\pd{\tilde{\mathcal{E}}_{\mathbf{k},{\theta}}}{k_{\mu}} + \frac{e}{\hbar} \tilde{\Omega}^{\mu \nu} E_\nu\right.\\
  +&\frac{e^2}{\hbar^2} \left( \Omega^{\mu \nu} B_{ \nu \gamma} \Omega^{\gamma \delta} E_\delta +  \frac{1}{2} {\Omega}^{\delta \gamma} B_{\delta \gamma} {\Omega}^{\mu \nu} E_\nu  \right) 
\nonumber \\ 
\nonumber
 +& \left.\frac{e}{\hbar^2} \left( \tilde{\Omega}^{\mu \nu} B_{ \nu \gamma} \pd{\tilde{\mathcal{E}}_{\mathbf{k},{\theta}}}{k_{\gamma}} + \frac{1}{2} \tilde{\Omega}^{\gamma \nu} B_{\gamma \nu} \pd{\tilde{\mathcal{E}}_{\mathbf{k},{\theta}}}{k_{\mu}} \right) 
  \right]f(\tilde{\mathcal{E}}_{\mathbf{k},{\theta}},\mu) \\
  +&\cdots \,.
  \label{eq:current}
 \end{align}
We are interested in the spatial components of current density $j^{i}$ generated when $\mathbf{E} = 0$. 
Keeping terms potentially linear in $\mathbf{B}$ results in~\footnote{We note that to rewrite the term in the second line of Eq.~\eqref{eq:current2}, which corresponds to $j^{i}_{b}$, we have used that only $E_{\theta}$ and $B_{ij} = \epsilon_{ijk} B_k$ are non-zero. This allows us to fully antisymmetrize the expression, $\epsilon_{\mu\nu\gamma\delta}$ guarantees that exactly one index is $\theta$.}
\resizebox{\linewidth}{!}{%
\begin{minipage}{\linewidth}
\begin{align}
\nonumber
j^i =&  e \int_{\mathbb{T}^3} \frac{ \text{d}^3 k }{(2\pi)^3} 
  \left[ \left( \frac{1}{\hbar} \frac{\partial \tilde{\mathcal{E}}}{\partial k_i }+  \dot{{\theta}}  \tilde{\Omega}^{i{\theta}}\right) - 
\frac{1}{8}\frac{e}{\hbar}\left(\epsilon_{\mu\nu\gamma\delta} \Omega^{\mu\nu}\Omega^{\gamma\delta}\right) \dot{{\theta}}B_{i}\right.\\
&\left.
+  \frac{1}{2}\frac{e}{\hbar^2}\left( \epsilon_{lmn}
\tilde{\Omega}^{lm} \frac{\partial  \tilde{\mathcal{E}}}{\partial k_n} \right)
  B_{i}\right]f(\tilde{\mathcal{E}}_{\mathbf{k},{\theta}},\mu) +\cdots\,,
 \label{eq:current2}
 \end{align}
 \vspace{0cm}
 \end{minipage}%
}
which is of the form $j^i = j^{i}_a+j^i_b+j^{i}_c$.
The last term, $j^i_c$ can be proven to be zero (see Supplemental Material) 
which is consistent with the absence of the chiral magnetic effect in the static limit~\cite{ZMS16,MP15}.
In the second term, $j^i_b$, keeping only linear order corrections in $B_{i}$ allows us to evaluate the distribution function at the unperturbed 
energy $\varepsilon_{\mathbf{k},\theta}$ leading to %
\begin{align}
j^i_{b} =& - \frac{1}{8}\frac{e^2}{\hbar} \int_{\mathbb{T}^3} \frac{ \text{d}^3 k }{(2\pi)^3} 
  \left[
\left( \epsilon_{\mu\nu\gamma\delta} \Omega^{\mu\nu}\Omega^{\gamma\delta}\right) \dot{{\theta}}  \right]f(\varepsilon_{\mathbf{k},{\theta}},\mu)B_{i} \,,
 \label{eq:currentjb}
 \end{align}
which we note is linear in magnetic field as desired and explicitly gauge invariant. 
To simplify $j^i_a$, we can expand the Fermi-Dirac distribution around its unperturbed  form $ f(\varepsilon_{\mathbf{k},{\theta}},\mu)$ %
 \begin{equation}
 \label{eq:exp}
 f(\tilde{\mathcal{E}}_{\mathbf{k},{\theta}},\mu) \sim f(\varepsilon_{\mathbf{k},{\theta}},\mu)
 +\dfrac{\partial f(\mathcal{E})}{\partial \mathcal{E}}\Big|_{\varepsilon_{\mathbf{k},{\theta}}}\tilde{\mathcal{E}'}+\cdots,
 \end{equation} 
where  $\tilde{\mathcal{E}'}=-\mathbf{m}_\mathbf{k,{\theta}}\cdot \mathbf{B}$. We obtain
\begin{align}
\nonumber
j^i_a =& 
e \int_{\mathbb{T}^3} \frac{ \text{d}^3 k }{(2\pi)^3}\left[  \dot{{\theta}}  \Omega^{i {\theta}} f(\varepsilon_{\mathbf{k},{\theta}},\mu) + 
\dot{{\theta}} \Omega_1^{i {\theta}} f(\varepsilon_{\mathbf{k},{\theta}},\mu)\right.\\
&\left.-\dot{{\theta}}  \Omega^{i {\theta}} \dfrac{\partial f(\mathcal{E})}{\partial \mathcal{E}}\Big|_{\varepsilon_{\mathbf{k},{\theta}}} \mathbf{m}_\mathbf{k,{\theta}}\cdot \mathbf{B}\right]
 \label{eq:currentja},
\end{align}
using that the integral of the Fermi velocity over the Fermi sea vanishes. 
The correction $\Omega^{i\theta}_1$ to the Berry curvature results from interband mixing and vanishes as $1/\Delta^{3}$ where $\Delta$ is the separation between different bands~\cite{GYS14}.
Taking $\Delta$ to be large, the low temperature limit and recasting the last term in \eqref{eq:currentja} as a Fermi surface contribution,  
the final response, which is the central result of this Letter, is given by
\begin{eqnarray}
 \label{eq:current}
j^i &=& \beta^{i} \dot{{\theta}}+(\alpha_{1}\delta^{ij}+\alpha^{ij}_{2})\dot{{\theta}}B_{j}+\mathcal{O}(1/\Delta^3) ,\\
 \label{eq:beta}
\beta^{i} &=& e \int_{\text{occ.}} \frac{ \text{d}^3 k }{(2\pi)^3}\Omega^{i {\theta}} ,\\
 \label{eq:alpha1}
\alpha_{1}&=& - \frac{1}{8}\frac{e^2}{\hbar}\int_{\text{occ.}} \frac{ \text{d}^3 k }{(2\pi)^3} 
\epsilon_{\mu\nu\gamma\delta} \Omega^{\mu\nu}\Omega^{\gamma\delta} ,\\
 \label{eq:alpha2}
\alpha^{ij}_{2}&=& -\frac{e}{\hbar} \int_{\mathrm{FS}} \frac{ \text{d}^2 k }{(2\pi)^3} \frac{1}{\left|\mathbf{v}_{\mathbf{k}}\right|} \Omega^{i {\theta}}_{\mathbf{k}} m^{j}_\mathbf{k,{\theta}},
\end{eqnarray}
where $\hbar \left|\mathbf{v}_{\mathbf{k}}\right| = \left| \partial\varepsilon_{\mathbf{k},{\theta}}/\partial\mathbf{k}\right|$.

The first term $\beta^i$ is independent of the magnetic field and captures the piezoelectric effect~\cite{vanderbiltpiezo} when ${\theta}$ corresponds to strain. 
For metals, the bulk current arises from the change in polarization involving occupied states.

The second term, $\alpha_1$, is the analogue of the isotropic magnetoelectric effect in insulators. 
Recall that in an insulating system a polarization in response to a static magnetic field is characterized by the momentum integral of a Chern-Simons three-form
determined by the band structure~\cite{qilong,essinmoorevanderbilt,Essin2010,malashevich}.
For the case of metals we find that the change in polarization depends on the variation with respect to $\theta$ and is determined by the integral of the second Chern-form, $\epsilon_{\mu\nu\gamma\delta} \Tr \Omega^{\mu\nu}\Omega^{\gamma\delta}$ over occupied states. It is exactly the derivative of the Chern-Simons three-form with respect to $\theta$. 

Two important remarks are in order.
First, the semiclassical approach only incorporates single band effects and thus $\Omega^{\mu\nu}$ is an Abelian $U(1)$ curvature and we need not trace over its components. 
This yields an isotropic magnetoelectric effect in our semiclassical treatment, which neglects terms resulting from cross-gap contributions, which vanish as $1/\Delta$~\cite{Essin2010}.
Second, the current generated by finite deformations is well defined since it is the integral of the second Chern-form.
The Chern-Simons three-form is only gauge-invariant if integrated over a closed manifold, so it does not correspond to a measurable quantity in metals; the static polarization is ill defined in metals.

Finally, the third term, $\alpha^{ij}_2$, is a novel Fermi surface contribution that is unique to metals.
It is the correction to the piezoelectric response at linear order in the magnetic field due to the orbital moment of the Bloch states. 
In what follows, we estimate the magnitude of all three terms contributing to the current to find that the Fermi surface contribution dominates the response.
\paragraph{Experimental feasibility}
An estimate of the observability of the current in Eq.~\eqref{eq:current} 
relies on the magnitude of the Berry curvature  $\Omega^{\mu\nu}$, which is common to all its terms.
We have distinguished two contributions to $\Omega^{\mu\nu}$ of distinct physical origin: the purely spatial part $\Omega^{ij}$ and the mixed $\Omega^{i{\theta}}$ terms.
The former defines the Hall conductivity $\sigma_{ij}=C_{ij}e^2/h$ in the $(i,j)$ plane through the Chern number $C_{ij} = \frac{1}{2\pi}\int d^2 k \Omega^{ij}$.
Since $C_{ij}$ is of the order of unity~\cite{Chang2013} or higher~\cite{berneviglargechern,zhanghighchern},
we expect $\Omega^{ij}\gtrsim \frac{a^2}{2\pi}$  where we estimate the cross sectional area of the unit cell in the $(i,j)$ plane using the lattice spacing $a$. 
To estimate $\Omega^{i{\theta}}$ we use previously known facts about the piezoelectric effect.
Identifying ${\theta}$ with a specific strain component $\epsilon^{jk}$ (${\theta} = \epsilon^{jk}$), the piezoelectric constant reads~\cite{Resta2007}
\begin{equation}
\beta^i_{jk} = \frac{\partial P^i}{\partial \epsilon^{jk}} = -e \int_{\text{occ.}} \frac{d^3 k}{(2\pi)^3} \Omega^{i{\theta}}.
\end{equation}
This formula only contains the electronic (clamped ion) contribution to the polarization response, typically smaller than the dominant contribution from the rearrangement of the ions. The electronic contribution can nonetheless be accessed independently in \emph{ab initio} calculations that estimate $\beta^i \sim \SI[per-mode=symbol]{1}{\coulomb\per\square\metre}$ \cite{Szabo1998} (suppressing the strain component indices for clarity).
It follows that $\Omega^{i{\theta}}\sim \beta^i \ \frac{a^3}{e}$
using the inverse cube of the lattice spacing as an approximate volume of the Fermi sea.

From the above estimate of the piezoelectric effect we can now approximate the magnitude of the remaining terms in Eq.~\eqref{eq:current},
$\alpha_{1}$ and $\alpha^{ij}_{2}$ given by Eqs.~\eqref{eq:alpha1} and \eqref{eq:alpha2} respectively, that are novel to this Letter.
The magnitude of the Fermi sea contribution $\alpha_1$ amounts to
\begin{equation}
\alpha_1 \sim \frac{e}{\hbar} a^2 C_{ij}\beta^k ,
\end{equation}
for a particular set of $i \neq j \neq k$ and neglecting the order one factor arising from the 
difference between a Fermi sea integral and a Brillouin zone integral.
Inserting $\beta^k\sim \SI[per-mode=symbol]{1}{\coulomb\per\square\metre}$, $a \sim 10^{-10} \si{\metre}$, $C_{ij} = 1$ we get $\alpha_1 \sim 10^{-5} \si[per-mode=fraction]{\ampere\second\per\tesla\per\square\meter}$. 

The estimate of the magnitude of the Fermi surface term $\alpha_2$, unique to metals, requires the magnitude of the orbital magnetic moment $|\mathbf{m}|$. A conservative estimate results in $|\mathbf{m}| \sim \mu_B \sim 10^{-23} \si[per-mode=symbol]{\joule\per\tesla}$ where $\mu_{B}$ is the Bohr magneton but it can be as large as $|\mathbf{m}|\sim 30\mu_{B}$ \cite{Xiao2010}. 
The area of the Fermi surface can be estimated as $1/a^{2}$, the cross section of the BZ which is the inverse of the cross section of the real space unit cell.
Taking $v_{F} \sim 10^6 \si[per-mode=symbol]{\metre\per\second}$, which is typical for metals but can be significantly smaller for lightly doped insulators near the band bottom,
and using our above estimate for $\Omega^{i\theta}$ we obtain
\begin{equation}
\alpha_2 \sim \frac{\beta^{i} m^{j} a}{\hbar v_F} \sim 10^{-4} \si[per-mode=fraction]{\ampere\second\per\tesla\per\square\meter}.
\end{equation}
Therefore we conclude that $\alpha_{2}\gtrsim\alpha_{1}$, and the Fermi surface contribution specific to metallic systems is dominant.

In addition, it is relevant to emphasize the following important points regarding experimental detection. 
Firstly strain rates at the order of $10^{-2}\si{\per\second}$ are achievable in the elastic regime using ultrasonic techniques~\cite{Langenecker1966,Green1975}.
For a sample at the cm scale, with cross sectional area $A_s \sim 10^{-4} \si{\square\metre}$ and a magnetic field of $1\si{\tesla}$~
\footnote{The semiclassical approach is applicable if $\hbar \omega_c = \hbar e B /m \ll \min\left( h / \tau_{MF}, k_B T \right)$ where $\tau_{MF} = l/v_F$ is the mean free time between scattering events and $l$ is the average distance between impurities. For $B=1\si{\tesla}$ the Landau level splitting is $\hbar\omega_{c}\sim7.27\times 10^{-4}$eV, a broadening similar to that achieved at a temperature of $T=8.44$K. Assuming a relatively pure material with 
$l \sim 10^{-7}\si{\meter}$ with $v_F \sim 10^6$m/s, the broadening from scattering is $h/\tau_{MF} \sim 10^{-2}$eV, which is two orders of magnitude larger than $\hbar\omega_c$, establishing the applicability of our approach.},
the current signal is of the order of $I_i = A_s j_i \sim 100~\si{\pico\ampere}$. Conventional ammeters have sensitivity extending to the pA range that is further improved in superconducting quantum interference devices.

Second, the magnetopiezoelectric effect is expected to coexist with the piezoelectric contribution, so accurate measurements over a range of magnetic fields are necessary for its detection. In our estimates $\alpha_{2}$ is proportional to and much smaller than $\beta$. However, $\beta$ gets contributions from the entire Fermi sea, while $\alpha_2$ only depends on Fermi surface properties. This allows suppression of $\beta$ without changing $\alpha_2$ in appropriately engineered band structures.

Third, the movement of the ions and the polarization of electrons in the valence bands induces a bound surface charge density. Part of the bulk current can be trapped screening it, possibly preventing its detection in our proposed setup (Fig.~\ref{fig:MPE}), but there is no reason to expect full cancellation. We note as well that pumping DC current is also possible by out-of-phase modulation of different strain components. Such a deformation path encircles a finite area in parameter space; the integral of the current for a pumping cycle is in general a non-vanishing, non-quantized value.
 
Finally, from the materials perspective we find that MnSi satisfies most requirements for these effects to manifest. 
It is a magnetically ordered, inversion breaking metal with complex Berry curvature patterns in the conduction bands that is very susceptible to strain~\cite{Binz2009,Lee2009,Faak2005,Shibata2015}. The magnetic order, however, is incommensurate and very sensitive to external magnetic field. An ideal candidate material would have simple easy axis ferromagnetic or N{\'e}el order that has vanishing susceptibility for magnetic fields in the ordering direction in the low temperature limit. 
The recently studied polar metals~\cite{SGW13,rondinelli}, while non-magnetic, would provide a platform to realize the field-independent piezoelectric response $\beta$.
Cold-atomic systems also offer an alternative; the current is related to an easily accesible observable, the center-of-mass velocity $\mathbf{v}_{c.m.}$ through $\mathbf{v}_{c.m.}=\mathbf{j}/n$ where $n$ is the density of the atomic cloud. Recently, $\mathbf{v}_{c.m.}$ has been exploited as a probe of topological properties~\cite{PZO16} and it is therefore plausible that the effects we discuss here can be observed in these systems as well.

 \paragraph{Conclusion}

We have calculated a novel magnetopiezoelectric response in inversion and time-reversal breaking metals subjected to static magnetic field and dynamic strain. Similar to the anomalous Hall effect in metals which can be viewed as a generalization of the quantized anomalous Hall effect in insulators, our results for magnetopiezoelectricity generalize the magnetoelectric response of insulators to metals. As a key difference, we find an additional Fermi surface contribution that relies on a finite orbital moment of the electrons, that is unique to metals and likely dominates the effect in real systems.

 \paragraph{Acknowledgements}
 
We are grateful to Fernando de Juan and Hannah Price for enlightening discussions and correspondence. 
We acknowledge financial support from the Marie Curie Programme under EC Grant agreement No.~653846. (A.G.G.), NSF Grant No.~DMR-1507141 (D.V.), AFOSR MURI (R.I.), and the Quantum Materials program of LBNL (J.E.M.).  JEM acknowledges travel support from the Simons Center for Geometry and Physics and a Simons Investigatorship.

 \bibliography{dynamics}
 
\appendix
\onecolumngrid
\section{Supplemental Material: Vanishing of $j^{i}_c$}
In this section we prove that $j^{i}_c$, defined as 
\begin{equation}
j^i_c=  e \int_{\mathbb{T}^3} \frac{ \text{d}^3 k }{(2\pi)^3} 
\frac{1}{2}\frac{e}{\hbar^2}\left( \epsilon_{lmn} \tilde{\Omega}^{lm} \frac{\partial  \tilde{\mathcal{E}}}{\partial k_n} \right) B_{i}f(\tilde{\mathcal{E}}_{\mathbf{k},{\theta}},\mu)\,,
 \label{eq:jic}
\end{equation}
in the main text is zero. 
Integrating by parts we obtain
\begin{align}
j^{i}_c= & \frac{1}{2}\frac{e}{\hbar^2}   \int_{\mathbb{T}^3} \frac{ \text{d}^3 k }{(2\pi)^3} \left[ 
-\tilde{\mathcal{E}} \left( 
\epsilon_{lmn}\frac{\partial \tilde{\Omega}^{lm}}{\partial k_n }\right)
  \right.f(\tilde{\mathcal{E}}_{\mathbf{k},{\theta}},\mu)
    - \left.
\tilde{\mathcal{E}} \left( \epsilon_{lmn}
\tilde{\Omega}^{lm}\frac{\partial f(\tilde{\mathcal{E}}_{\mathbf{k},{\theta}},\mu)}{\partial k_n }\right)
  \right]B_{i}
  \label{eq:jc}
 \end{align}
 The first term is the divergence of the Berry field strength and vanishes.
 The second term in the low temperature limit can be rewritten as the total the Berry flux over all Fermi surfaces, which vanishes as well. 
%
This results is consistent with the absence of the chiral magnetic effect in the static limit~\cite{ZMS16,MP15}.%

\end{document}